\documentclass{PoS}
 
\def\Pom{I\!\!P}
\def\Reg{I\!\!R}

\title{Applications of the tensor pomeron model\\
to exclusive central diffractive meson production}

\ShortTitle{Applications of the tensor pomeron model to exclusive central diffractive meson production} 

\author{\speaker{Piotr LEBIEDOWICZ}\\
Institute of Nuclear Physics PAN, PL-31-342 Cracow, Poland\\
E-mail: \email{Piotr.Lebiedowicz@ifj.edu.pl}} 

\author{Otto NACHTMANN\\
Institut f\"ur Theoretische Physik, Universit\"at Heidelberg, Philosophenweg 16, D-69120 Heidelberg, Germany\\
E-mail: \email{O.Nachtmann@thphys.uni-heidelberg.de}}

\author{Antoni SZCZUREK\\
Institute of Nuclear Physics PAN, PL-31-342 Cracow, Poland, and\\
University of Rzesz\'ow, PL-35-959 Rzesz\'ow, Poland\\
E-mail: \email{Antoni.Szczurek@ifj.edu.pl}}

\abstract{
We discuss exclusive central diffractive production of scalar 
($f_{0}(980)$, $f_{0}(1370)$, $f_{0}(1500)$), 
pseudoscalar ($\eta$, $\eta'(958)$), 
and vector ($\rho^{0}$) mesons in proton-proton collisions.
We show that high-energy central production of mesons
could provide crucial information on the spin structure of the soft pomeron.
The amplitudes are formulated in terms of
effective vertices respecting standard rules of Quantum Field Theory
and propagators for the exchanged pomeron and reggeons.
For the scalar and pseudoscalar meson production, 
in most cases, two lowest orbital angular momentum - spin couplings 
are necessary to describe WA102 experimental differential distributions.
Different pomeron-pomeron-meson tensorial (vectorial)
coupling structures are possible in general.
For the $\rho^{0}$ production the photon-tensor pomeron/reggeon exchanges 
are considered
and the coupling parameters are fixed from the H1 and ZEUS experimental data
of the $\gamma p \to \rho^{0} p$ reaction.
We present first predictions of this mechanism
for the $pp \to pp (\rho^{0} \to \pi^{+} \pi^{-})$ reaction being
studied at COMPASS, RHIC, Tevatron, and LHC.
We analyse influence of the experimental cuts on integrated cross section
and various differential distributions for produced mesons.
}

\FullConference{XXII. International Workshop on Deep-Inelastic Scattering and Related Subjects \\
          28 April - 2 May 2014 \\
          Warsaw, Poland} 

\begin{document}

\section{Introduction}

\begin{figure} 
\centering
\includegraphics[width=0.28\textwidth]{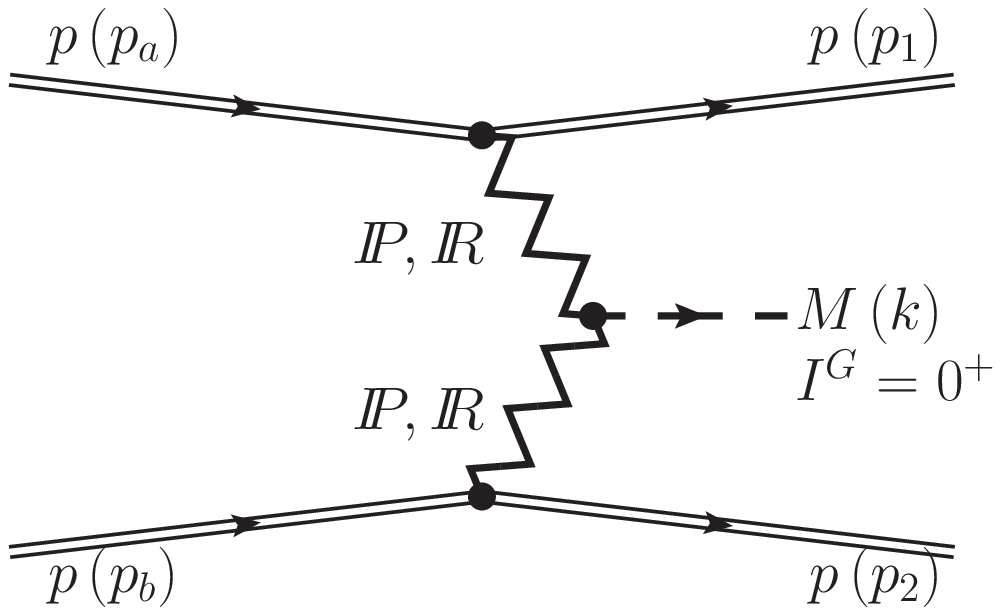}
\qquad
\includegraphics[width=0.25\textwidth]{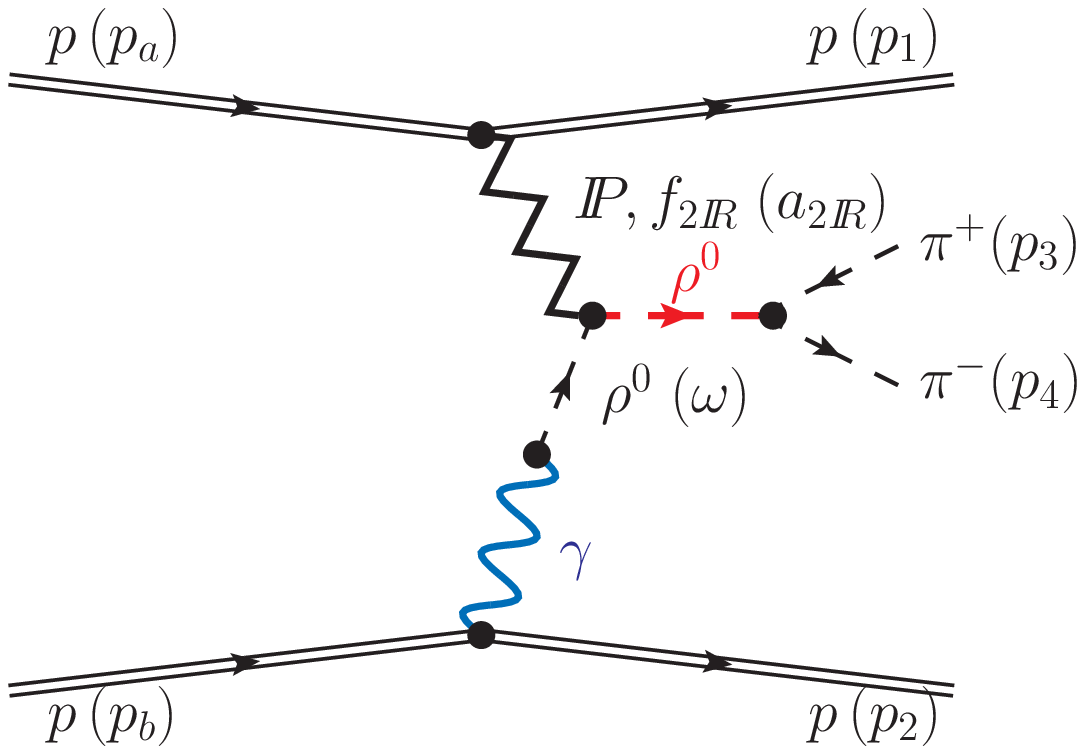}
\caption{
Left panel: Diagram of the central exclusive $I^{G} = 0^{+}$
mesons production by a fusion of pomeron-pomeron, 
pomeron-reggeon, reggeon-pomeron, or reggeon-reggeon.
Right panel: Diagram of the central exclusive $\rho^{0}$ meson production 
via pomeron/reggeon-photon (or photon-pomeron/reggeon) exchanges
and its subsequent decay into $\pi^+ \pi^-$ pair.
}
\label{fig:0}
\end{figure}
There is a growing interest in understanding 
the mechanism of exclusive meson production
both at low and high energies, for a review see \cite{Lebiedowicz_Thesis}.
In Ref.~\cite{LNS14} we performed application of the "tensorial" pomeron model \cite{EMN14} 
for the exclusive production of scalar ($J^{PC} = 0^{++}$) 
and pseudoscalar ($J^{PC} = 0^{-+}$) mesons 
in the $pp \to pp M$ reaction 
and we compared results of our calculations with WA102 experimental data.
At high energies the dominant contribution to
this reaction comes from pomeron-pomeron (or pomeron-reggeon) fusion,
see Fig.~\ref{fig:0}~(left panel).
While it is clear that the effective Pomeron must be a colour singlet, 
its spin structure and couplings to hadrons are not finally established. 
Although, it is also commonly assumed in the literature that the pomeron has
a "vectorial" nature \cite{DDLN02}
there are strong hints from the Quantum Field Theory
that it has rather "tensorial" properties \cite{EMN14}.

Indeed, tests for the helicity structure of the pomeron 
have been devised in \cite{Arens:1996xw}
for diffractive contributions to electron-proton scattering, that is,
for virtual-photon--proton reactions.
For central meson production in proton-proton collisions such tests
were discussed in \cite{Close:1999is,Close:1999bi,Close:2000dx}.
The general structure of helicity amplitudes of the simple Regge behaviour
was also considered in Ref.~\cite{Kaidalov:2003fw,Petrov:2004hh}.
The detailed structure of the amplitudes
depends on dynamics and cannot be predicted from the general principles of Regge theory.

We focus on exclusive production of $\rho^0$ resonance \cite{LNS14_rho}
in the context of theoretical concept of tensor pomeron proposed in Ref.\cite{EMN14}. 
The diagrams to be considered are shown in Fig.~\ref{fig:0}~(right panel).
Needless to say that at lower energies there
is also an important $\rho^0 \to \pi^+ \pi^-$ background from
the non-central diffractive processes.
\footnote{There the dominant mechanism is 
the hadronic bremsstrahlung-type mechanism. 
Similar processes were discussed at high energies
for the exclusive reactions:
$pp \to nn \pi^{+}\pi^{+}$ \cite{LS11},
$p p \to p p \pi^0$ \cite{LS_pi0},
$p p \to p p \omega$ \cite{CLSS},
$p p \to p p \gamma$ \cite{LS13}.}
Two of us proposed some time ago a simple Regge-like model for 
the $\pi^+ \pi^-$ and $K^{+} K^{-}$ continuum based on the exchange of two 
pomerons/reggeons \cite{LS10, LPS11, LS12, Lebiedowicz_Thesis}.

\section{Sketch of formalism}

In Ref.~\cite{LNS14} we discussed differences between results of the 
"tensorial pomeron" and "vectorial pomeron" models 
of central exclusive production for the scalar and pseudoscalar mesons.
In order to calculate these contributions we must know 
the effective $I\!\!P$ propagator, the $I\!\!P p p$ and $I\!\!P I\!\!P M$ vertices.
The formulae for corresponding propagators and vertices 
are presented and discussed in detail in Refs.~\cite{EMN14,LNS14}.
In \cite{LNS14}
we gave the values of the lowest orbital angular momentum $l$
and of the corresponding total spin $S$,
which can lead to the production of $M$ in the fictitious fusion of two
tensorial or vectorial "pomeron particles".
In most cases one has to add coherently amplitudes for two couplings 
with different orbital angular momentum and spin of two "pomeron particles". 
The corresponding coupling constants are not known 
and have been fitted to existing experimental data. 

Here we focus on exclusive central production of $\rho^0$ resonance.
Due to its quantum numbers this resonance state can be
produced only by photon-pomeron or photon-reggeon mechanisms, 
see Fig.~\ref{fig:0}~(right panel).
In the amplitude for the $\gamma p \to \rho^{0} p$ subprocess
we included both pomeron and $f_{2 \Reg}$ exchanges.
All effective vertices and propagators have been taken here
from Ref.~\cite{EMN14}.
The $\Pom \rho \rho$ vertex is given in \cite{EMN14} by formula (3.47)
and the propagator of the tensor-pomeron exchange by (3.10).
The decay vertex for $\rho^0 \to \pi^+ \pi^-$ is well known 
(e.g. see (3.35) of \cite{EMN14}).
All the details will be given in \cite{LNS14_rho}.

\section{Results}
\label{section:Results}

\begin{figure} 
\centering
\includegraphics[width=0.32\textwidth]{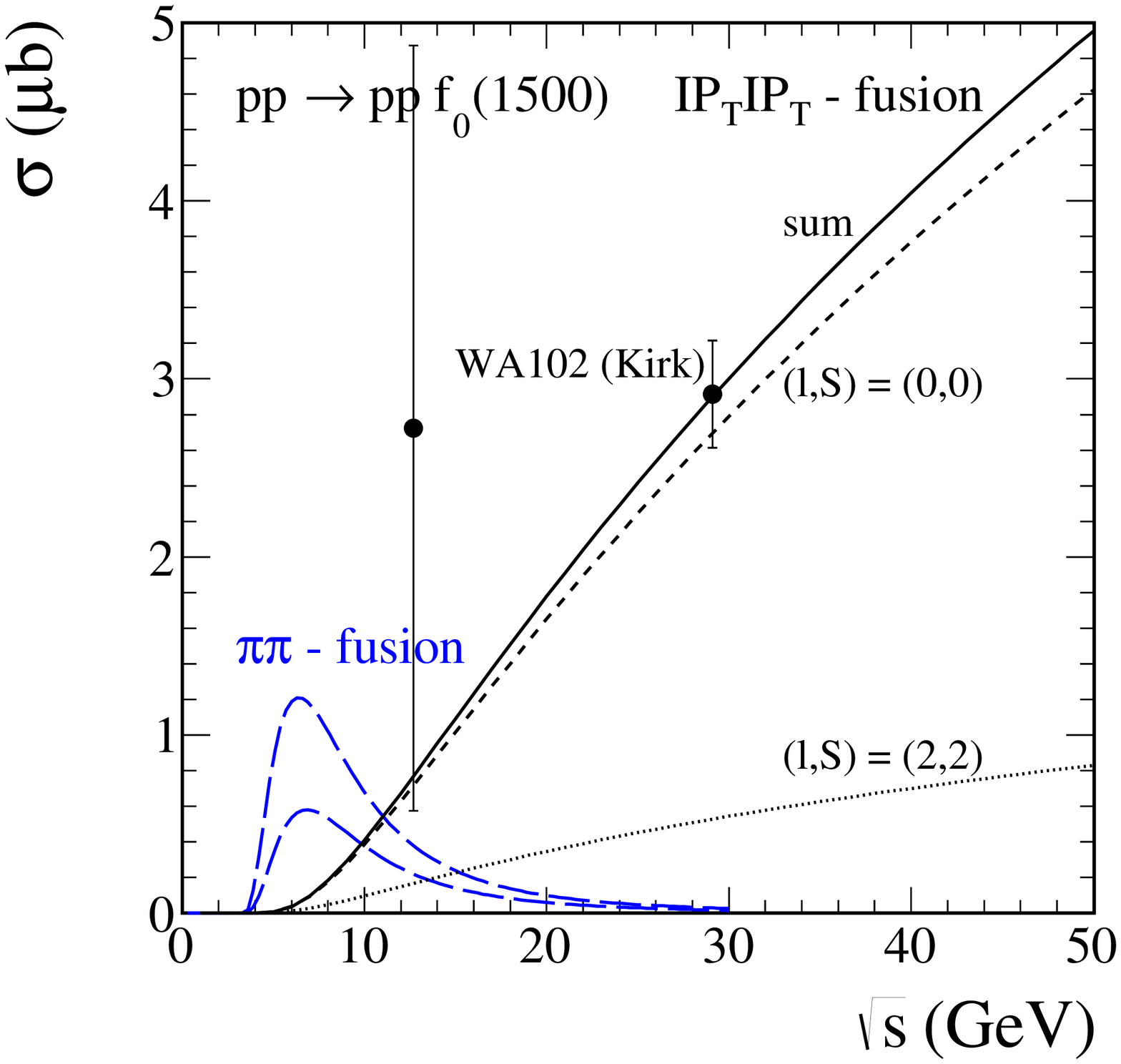}
\includegraphics[width=0.32\textwidth]{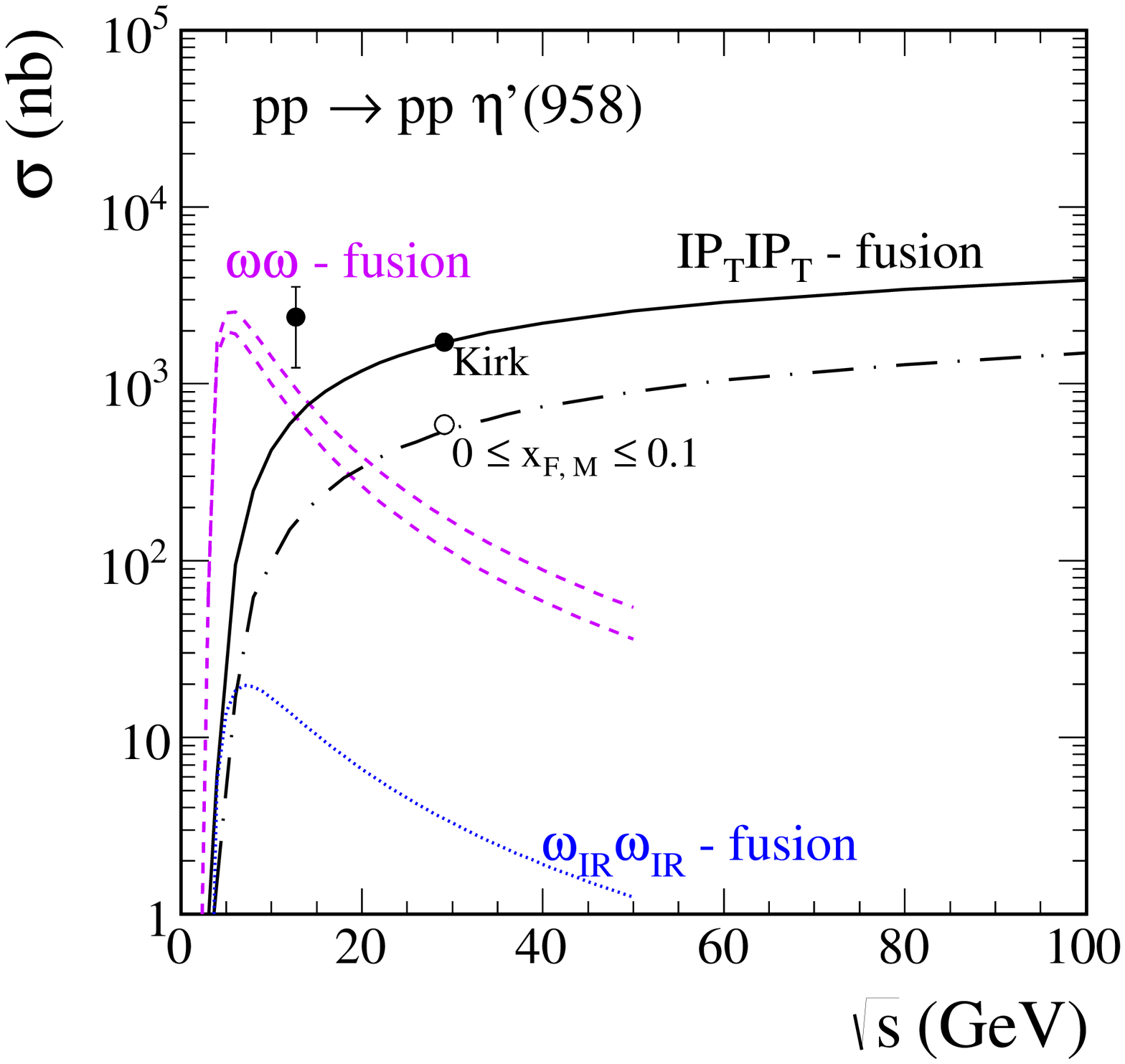}
\caption{
Cross section for the $pp \to pp f_{0}(1500)$ (left panel) 
and $pp \to pp \eta'(958)$ (right panel) reactions
as a function of proton-proton center-of-mass energy $\sqrt{s}$.
The experimental data are from the WA102 experiment 
at $\sqrt{s} = 29.1 $~GeV \cite{Kirk:2000ws},
and for the Feynman-$x_{F}$ interval $0 \leqslant x_{F,M} \leqslant 0.1$ \cite{Barberis:1998ax}.
The pomeron-pomeron fusion (the black solid lines) dominates at higher energies.
In the left panel we show the individual contributions to the cross section
with $(l,S) = (0,0)$ (short-dashed line) and $(l,S) = (2,2)$ (dotted line).
The $\pi\pi$-fusion contribution is important for the $f_{0}(1500)$ meson production
at lower energies
while the $\omega\omega$-fusion contribution for the $\eta'$ meson production.
}
\label{fig:1}
\end{figure}

The theoretical results are compared with the WA102 experimental data
in order to determine the model parameters.
In Fig.\ref{fig:1} for example we present our result for the integrated cross sections
of the exclusive $f_{0}(1500)$ (left panel) and $\eta'$ (right panel) meson production
as a function of centre-of-mass energy $\sqrt{s}$.
At high energies the dominant contribution
comes from pomeron-pomeron ($I\!\!P I\!\!P$) fusion;
see Fig.\ref{fig:0}~(left). 
Non-leading terms arise from
reggeon-pomeron ($I\!\!R I\!\!P$) and reggeon-reggeon ($I\!\!R I\!\!R$) exchanges.
We assume that the energy is high enough that we can consider
pomeron-pomeron-meson ($I\!\!P I\!\!P M$) fusion.
At low energy the $\pi\pi$-fusion contribution \cite{Szczurek:2009yk} dominates.
For pseudoscalar meson production we can expect large contributions from
$\omega \omega$ exchanges due to the large coupling of the $\omega$ meson to the nucleon.
At higher subsystem energies squared $s_{13}$ and $s_{23}$
the meson exchanges are corrected to obtain the high energy behaviour
appropriate for $\omega$-reggeon exchanges.
This seems to be the case for the $\eta$ meson production
where we have included also exchanges of subleading 
trajectories which improve the agreement with experimental data.

\begin{figure} 
\centering
\includegraphics[width=0.32\textwidth]{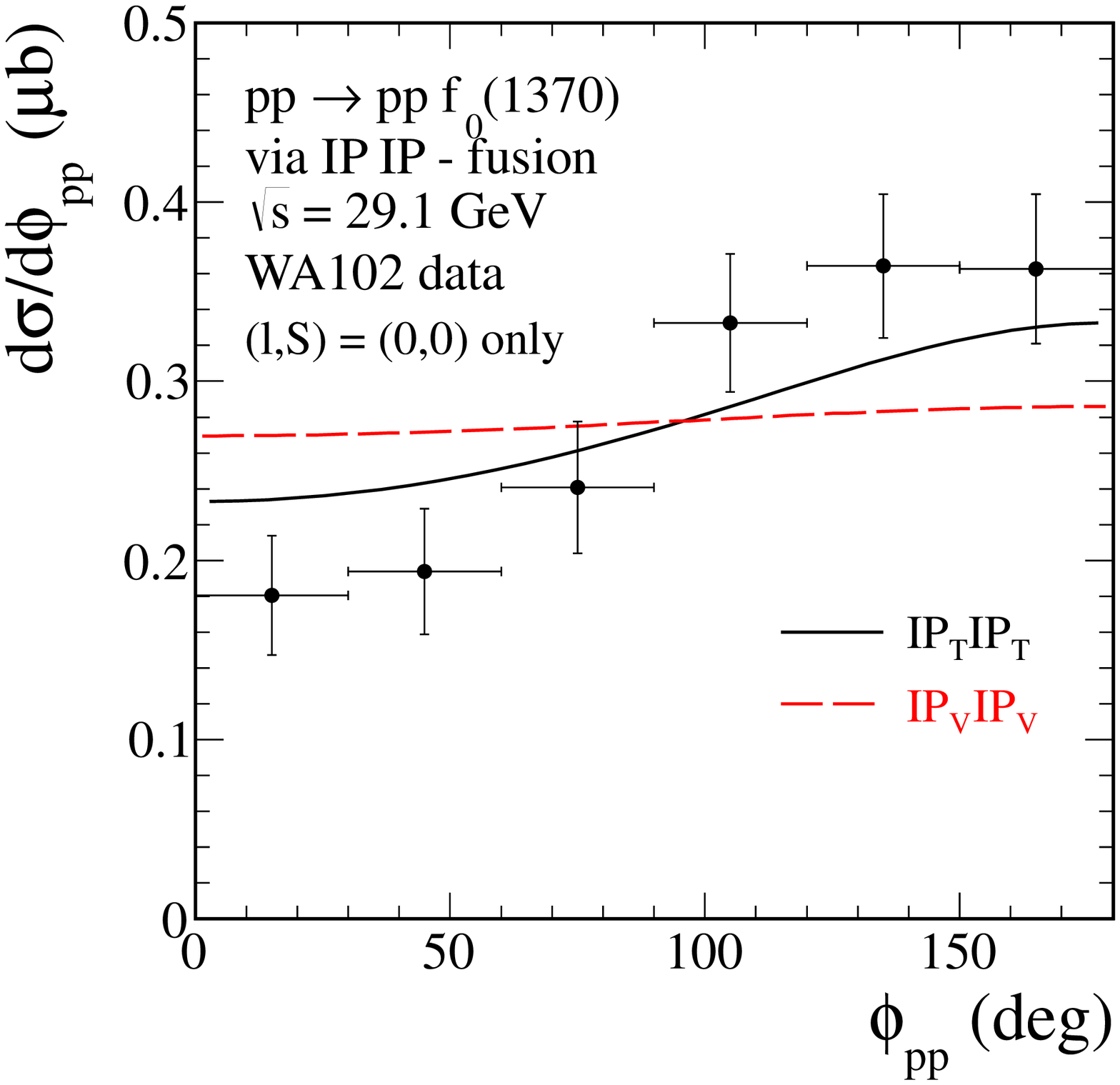}
\includegraphics[width=0.32\textwidth]{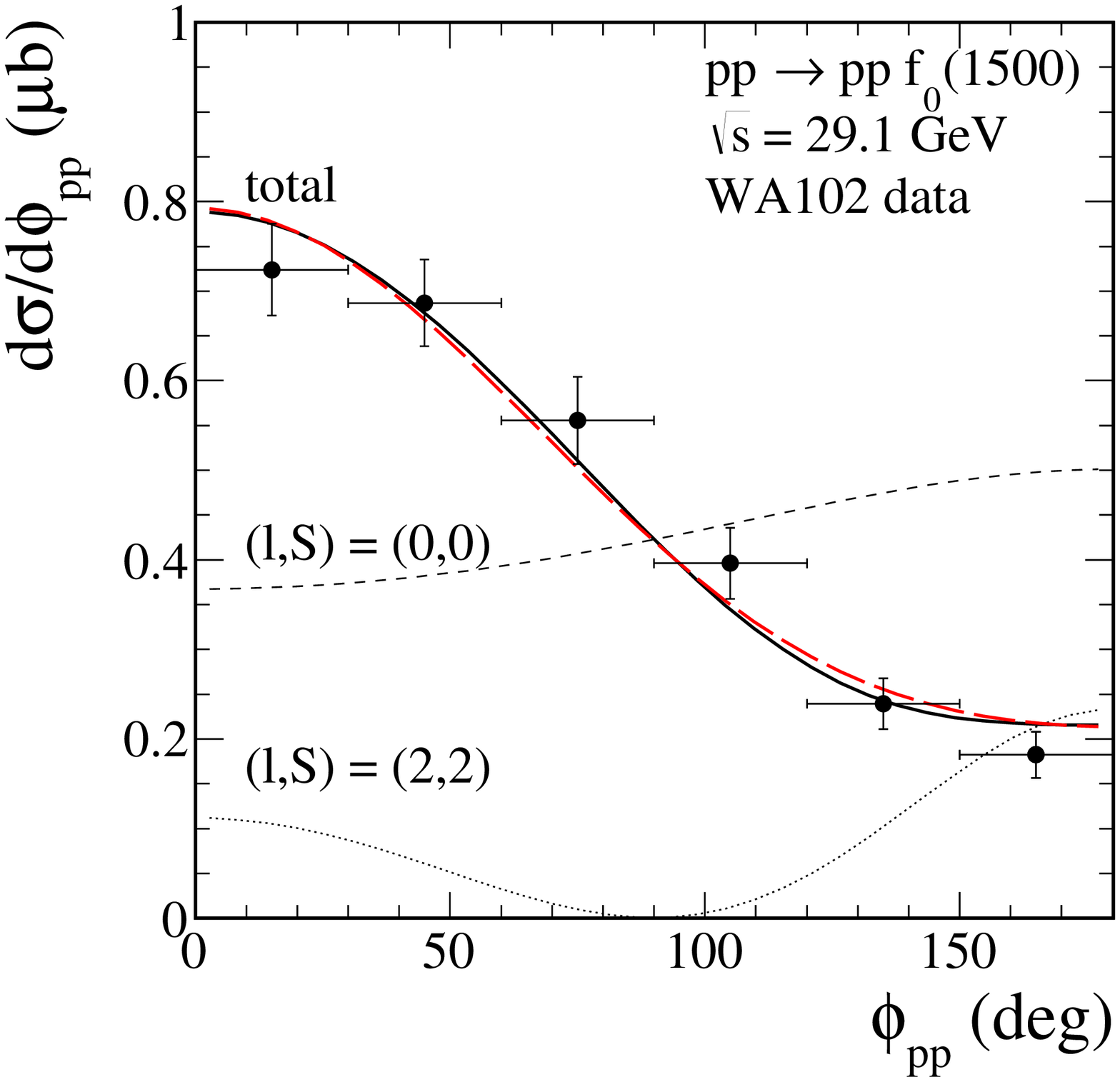}
\includegraphics[width=0.32\textwidth]{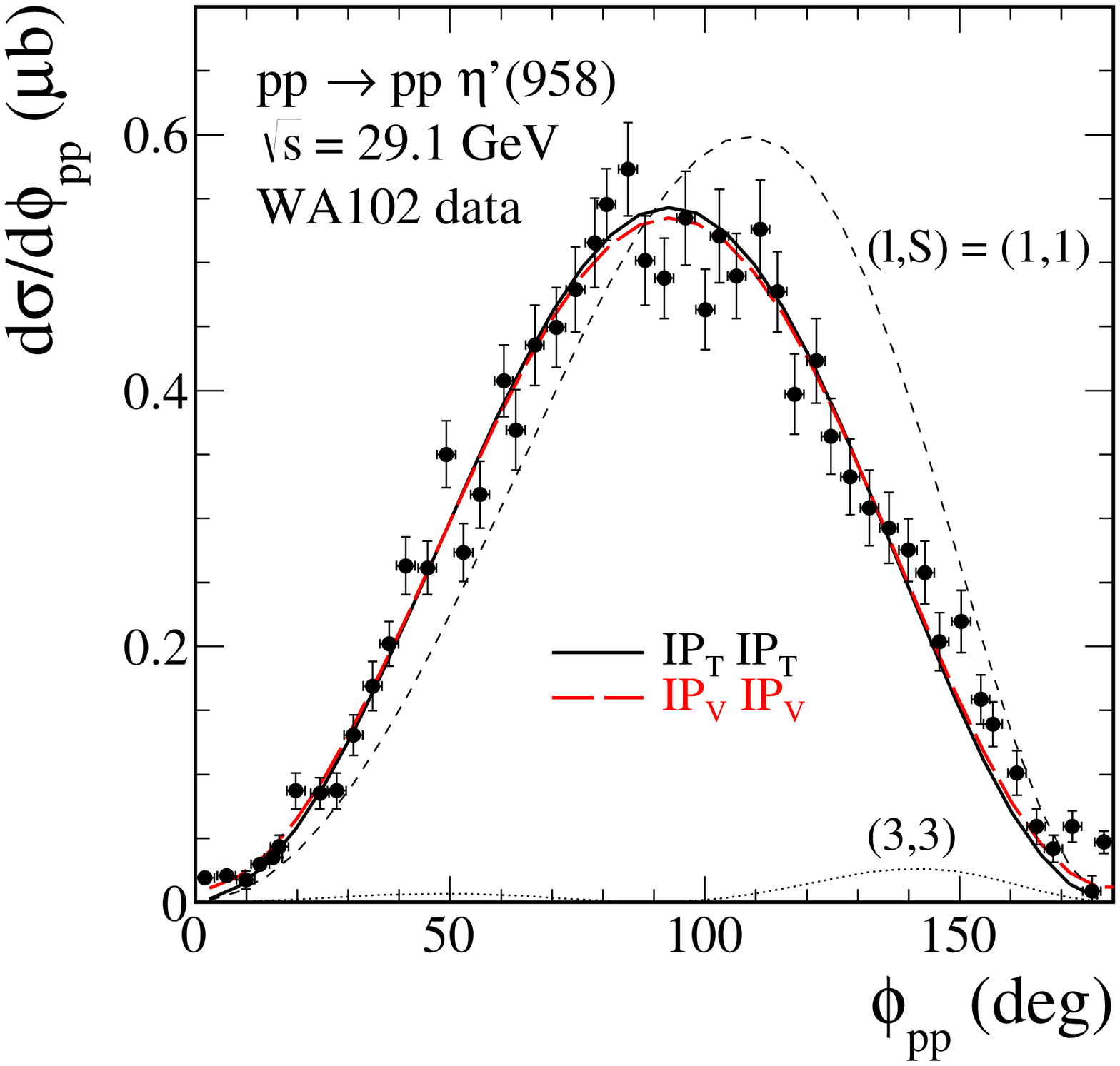}
\caption{
Distributions in azimuthal angle between outgoing protons
for the central exclusive $f_{0}(1370)$, $f_{0}(1500)$ and $\eta'(958)$ mesons production 
by a fusion of two tensor (the black solid line) 
or two vector (the red long-dashed line) pomerons.
The WA102 experimental data points from \cite{Barberis:1999cq}
have been normalized to the total cross section 
from \cite{Kirk:2000ws} at $\sqrt{s} = 29.1$~GeV.
The solid line is the result including coherently two $(l,S)$ couplings.
For the tensorial pomeron case we show the individual $(l,S)$ 
spin contributions to the cross section.
}
\label{fig:2}
\end{figure}

In Fig.\ref{fig:2} we show the distribution in azimuthal angle $\phi_{pp}$
between outgoing protons for central exclusive meson production
by the fusion of two tensor (solid line) or two vector (long-dashed line) pomerons 
at $\sqrt{s} = 29.1$~GeV.
The strengths of the $(l,S)$ couplings
were adjusted to roughly reproduce the WA102 data from \cite{Barberis:1998ax}.
The tensorial pomeron with the $(l,S) = (0,0)$
coupling alone already describes the azimuthal angular correlation
for $f_{0}(1370)$ meson reasonable well. 
The vectorial pomeron with the $(l,S) = (0,0)$ term alone is disfavoured here.
The preference of the $f_{0}(1370)$ for the $\phi_{pp} \approx \pi$ domain in contrast to
the enigmatic $f_{0}(980)$ and $f_{0}(1500)$ scalars has been observed by 
the WA102 Collaboration \cite{Barberis:1999cq}.
The contribution of the $(0,0)$ component alone
is not able to describe the azimuthal angular dependence for these states.
We observe a large interference of two $(l,S)$ components in the amplitude.
For production of pseudoscalar mesons in both pomeron models 
the theoretical distributions are somewhat skewed due to phase space angular dependence.
The contribution of the $(1,1)$ component alone in the tensorial pomeron model
is not able to describe the WA102 data \cite{Barberis:1998ax}.

%
%

\begin{figure} 
\centering
\includegraphics[width=0.32\textwidth]{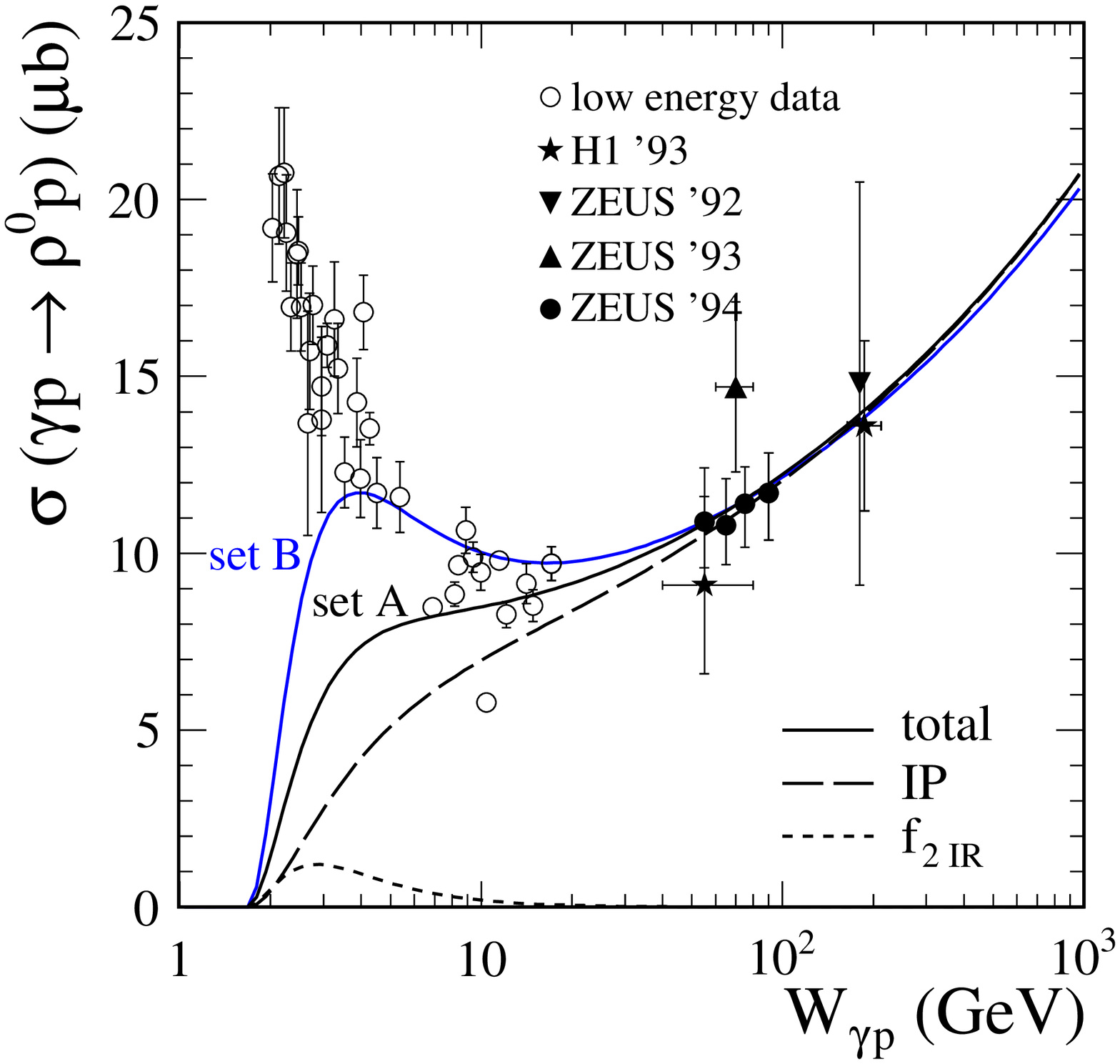}
\includegraphics[width=0.32\textwidth]{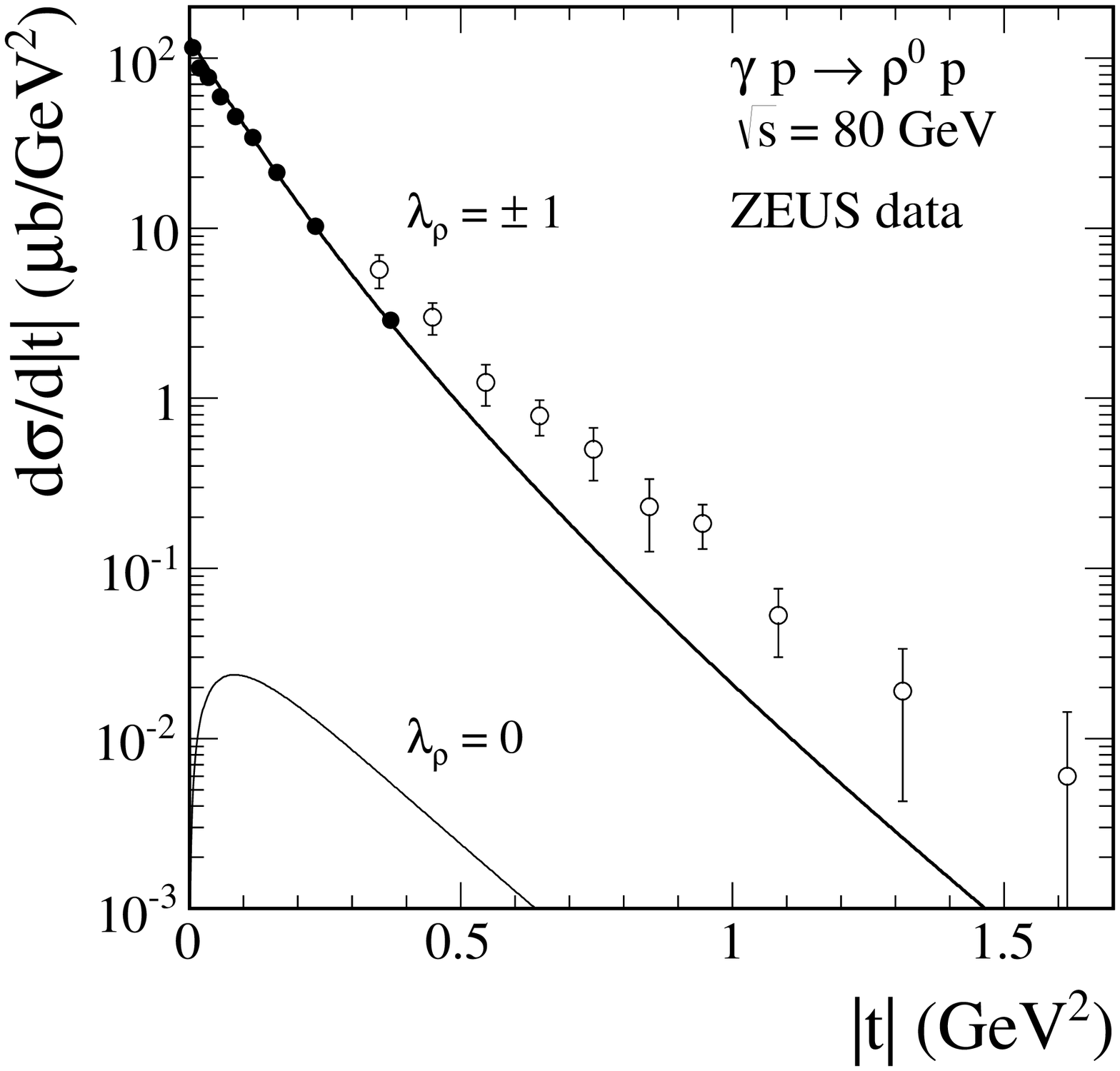}
\caption{
Left panel:
The elastic $\rho^{0}$ photoproduction cross section
as a function of center-of-mass energy.
Our results are compared with the low energy data and HERA data (solid marks),
see \cite{CLSS} for references.
The solid line correspond to results with both the pomeron and $f_{2 \Reg}$ exchanges.
The individual pomeron and reggeon exchange contributions
are denoted by the long-dashed and short-dashed lines, respectively.
The black lines represent parameter set~A of coupling constants
obtained for default values of $\Pom \rho \rho$ 
and $f_{2 \Reg} \rho \rho$ couplings
in both the $\Gamma^{(0)}$ and $\Gamma^{(2)}$ tensor functions \cite{EMN14}
while the blue line represents set~B for the $\Gamma^{(2)}$ function alone
and more accurately describes the low-energy experimental data
for a larger value of $b_{f_{2 \Reg} \rho \rho}$ coupling.
Right panel: 
The differential cross section $d\sigma/d|t|$ for the $\gamma p \to \rho^{0} p$ process.
The ZEUS data at low $|t|$ (at $\gamma p$ average center-of-mass 
energy $<\sqrt{s}> = 71.7$~GeV \cite{Breitweg:1997ed}
and at higher $|t|$ (at $<\sqrt{s}> = 94$~GeV) \cite{Breitweg:1999jy} are shown. 
The top and bottom lines represent the results at $\sqrt{s} = 80$~GeV
for $\rho^{0}$ meson transverse ($\lambda_{\rho} = \pm 1$)
and longitudinal ($\lambda_{\rho} = 0$) polarisation, respectively.
Here we used parameter set~A of coupling constants.
}
\label{fig:3}
\end{figure}

Fig.~\ref{fig:3} (left panel) shows 
the integrated cross section for the $\gamma p \to \rho^{0} p$ reaction
as a function of center-of-mass energy together with the experimental data.
In our calculations two parameter sets of coupling constants are used, see \cite{LNS14_rho}.
In the right panel we show the differential cross section 
for elastic $\rho^{0}$ photoproduction.
The calculations performed at $\sqrt{s} = 80$~GeV 
are compared with ZEUS data \cite{Breitweg:1997ed, Breitweg:1999jy}.
We can see that the amplitude for longitudinal $\rho^{0}$ meson polarisation
is negligible and vanishes at $t = 0$.

\begin{figure} 
\centering
\includegraphics[width=0.32\textwidth]{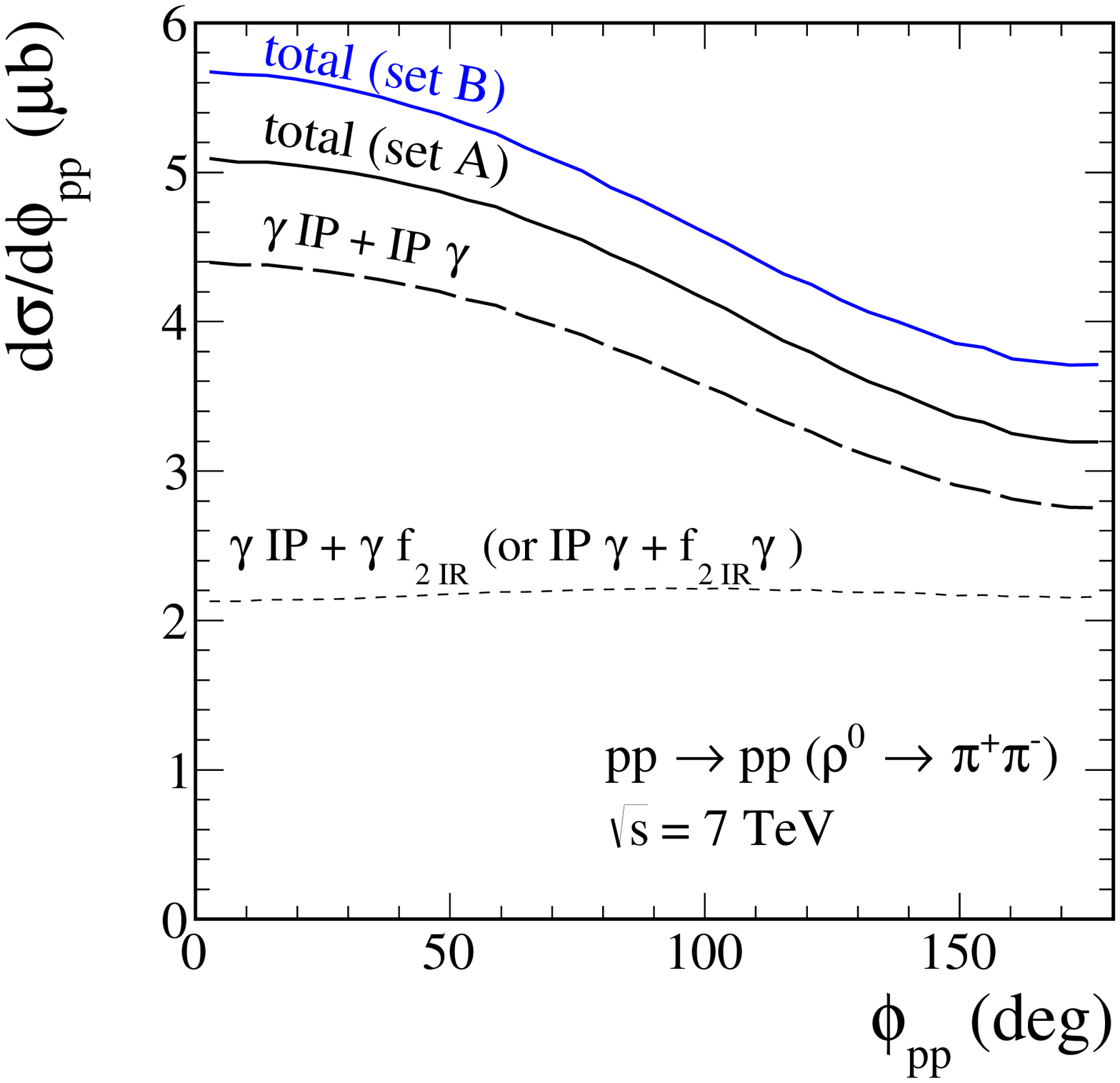}
\includegraphics[width=0.32\textwidth]{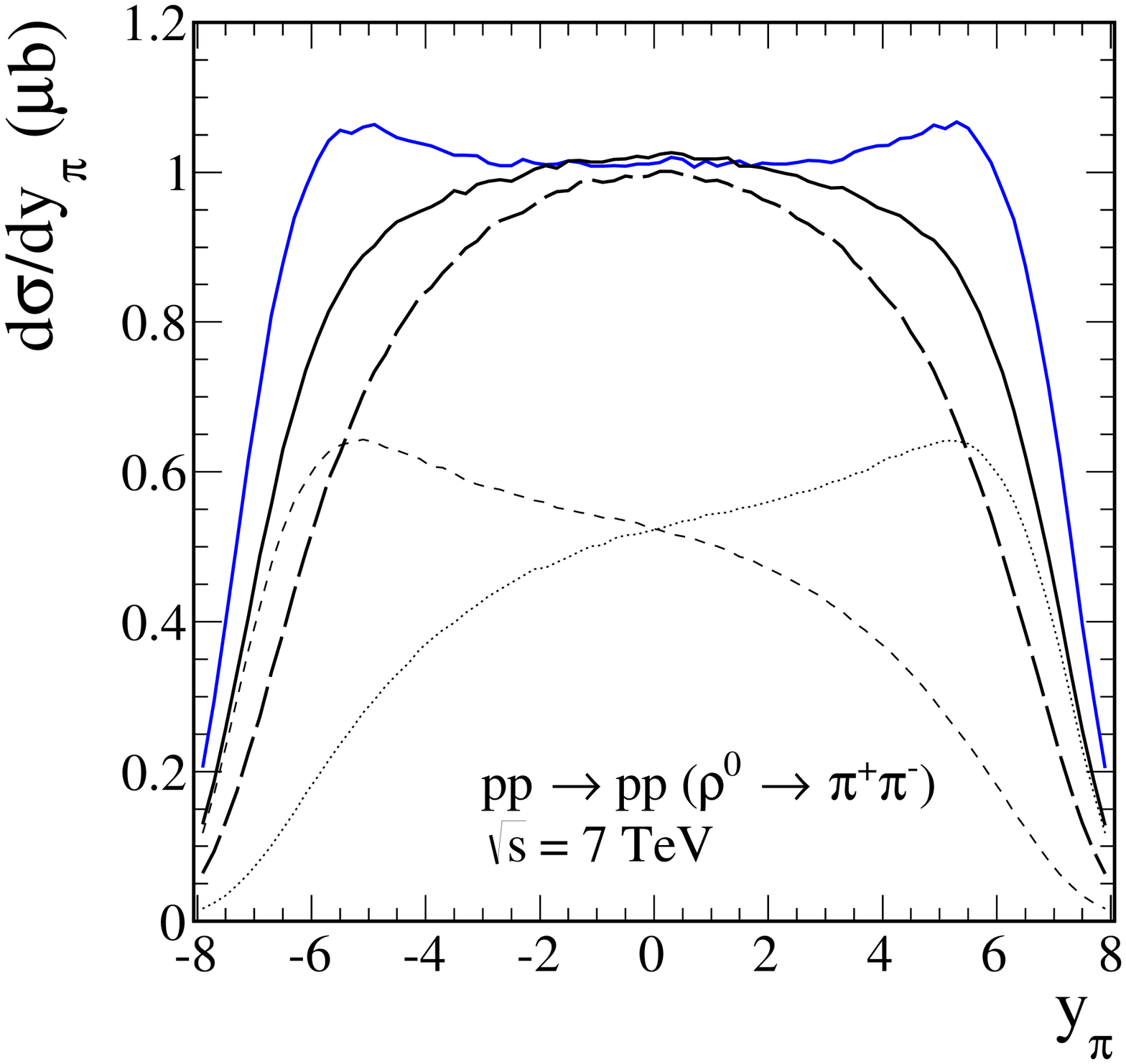}
\caption{
Distributions 
in azimuthal angle between outgoing protons (left panel)
and in the pion rapidity (right panel) at $\sqrt{s} = 7$~TeV.
The black (lower) and blue (upper) solid lines are obtained 
for parameter set~A and set~B of coupling constants, respectively,
and correspond to the situation when both
the pomeron and $f_{2 \Reg}$ exchanges in the amplitude are included.
The short-dashed and dotted line represent contribution separately
from the pomeron-photon and photon-pomeron diagrams, respectively.
The black long-dashed line corresponds to the pomeron exchange alone and parameter set~A.
}
\label{fig:4}
\end{figure}

In Fig.~\ref{fig:4} we show distributions
in $\phi_{pp}$ (left panel)
and in the pion rapidity (right panel) at $\sqrt{s} = 7$~TeV.
For the $\phi_{pp}$ distribution, the effect of deviation from a constant is due to
interference of photon-pomeron and pomeron-photon amplitudes.
Similar effect was discussed first in Ref.~\cite{Schafer:2007mm} for the exclusive
production of $J/\psi$ meson. These correlations are quite different than
those for double-pomeron mechanism \cite{LPS11, Lebiedowicz_Thesis} 
and could be therefore
used at least to partial separation of these two mechanisms.
\footnote{The $\rho^0$ contribution in the azimuthal angle region ($\phi_{pp} < \pi/2$) 
should be strongly enhanced
in comparison to fully diffractive mechanism including absorption effects
due to the $pp$-rescattering.}
The rapidities of the two pions are strongly correlated. 
The $f_{2 \Reg}$ exchange included in the amplitude 
contributes at backward/forward pion rapidities
and its contribution is non-negligible even at the LHC energies.
This is similar as for the double pomeron/reggeon-exchanges 
in the fully diffractive mechanism, see \cite{LS10, LPS11, Lebiedowicz_Thesis}.

\section{Conclusions}

We have analysed proton-proton collisions with the exclusive central production of
scalar and pseudoscalar mesons.
We have presented the predictions of two different models of the soft pomeron. 
The first one is the commonly used model with vectorial pomeron which is, 
however, difficult to be supported from a theoretical point of view.
The second one is a recently proposed model of tensorial pomeron \cite{EMN14}, which,
in our opinion, has better theoretical foundations.
We have performed calculations of several differential distributions.
We wish to emphasize that the tensorial pomeron can, at least, equally well describe
experimental data on the exclusive meson production discussed here as the less 
theoretically justified vectorial pomeron frequently used in the literature. 
The existing low-energy experimental data do not allow to
clearly distinguish between the two models as the presence of subleading
reggeon exchanges is at low energies very probable for many reactions. 
Production of $\eta'$ meson seems to be less affected by contributions from subleading exchanges.
Pseudoscalar meson production could be of particular interest
for testing the nature of the soft pomeron
since there the distribution in the azimuthal angle $\phi_{pp}$
between the two outgoing protons may contain, for the tensorial pomeron,
a term which is not possible for the vectorial pomeron, see \cite{LNS14}.
The models contain only a few free coupling parameters to be determined by experiment.
The hope is, of course, that future experiments will be able to select
soft pomeron model. In any case, our models should provide good ``targets''
for experimentalists to shoot at.
It would clearly be interesting to extend the studies of central meson production
in diffractive processes to higher energies.
For the resonances decaying e.g. into the $\pi \pi$ channel an interference
of the resonance signals with the two-pion continuum has to be included in addition. 
This requires a consistent model of the resonances and the non-resonant background.

We have made first estimates of the contribution of exclusive 
$\rho^0$ production to the $p p \to p p \pi^+ \pi^-$ reaction.
We have shown some differential distributions in pion rapidities
as well as some observables related to final state protons. 
We have obtained that the $\rho^0$ contribution
constitutes 10-20\% of the double-pomeron/reggeon contribution
calculated in a simple Regge-like model \cite{LS10, LPS11, Lebiedowicz_Thesis}. 
We expect that the exclusive $\rho^0$ photoproduction and 
its subsequent decay are the main source of $P$-wave in 
the $\pi^+ \pi^-$ channel in contrast to even waves 
populated by the double-pomeron processes.
Different dependence on proton transverse momenta and 
azimuthal angle correlations between outgoing protons
could be used to separate the $\rho^0$ contribution.
We therefore conclude that the measurement of forward/backward
protons is crucial in better understanding of the mechanism
of the $p p \to p p \pi^+ \pi^-$ reaction. 

Future experimental data on exclusive meson production at high
energies should thus provide good information on the spin structure of 
the pomeron and on its couplings to the nucleon and the mesons.
On the other hand, the low energy data could help in understanding the role of 
subleading trajectories. 
Several experimental groups, e.g. 
COMPASS \cite{COMPASS},
STAR \cite{Turnau_DIS2014}, 
CDF \cite{CDF},
ALICE \cite{Schicker:2012nn}, 
ATLAS+ALFA\cite{SLTCS11}, and CMS+TOTEM \cite{TOTEM}
have potential to make very significant contributions to this program aimed at
understanding the spin structure of the soft pomeron.

This work was partially supported by the Polish National Science Centre
on the basis of decisions DEC-2011/01/N/ST2/04116 and DEC-2013/08/T/ST2/00165.

\end{document}